\documentclass[nofootinbib,superscriptaddress]{revtex4-1}
\usepackage{epsfig}
\usepackage{amsmath}
\usepackage{amssymb}
\begin{document}


\title{Elastic Coulomb breakup of $^{34}$Na}


\author{G. Singh}
\email{gagandph@iitr.ac.in}
\affiliation{Department of Physics, Indian Institute of Technology - Roorkee, 247667, INDIA}
\author{Shubhchintak}
\email{shub.shubhchintak@tamuc.edu}
\affiliation{Department of Physics and Astronomy, Texas A$\&$M University - Commerce, 75429, USA}
\author{R. Chatterjee}
\email{rcfphfph@iitr.ac.in}
\affiliation{Department of Physics, Indian Institute of Technology - Roorkee, 247667, INDIA}


\date{\today}

\begin{abstract}
\begin{description}
\item[Background]
$^{34}$Na is conjectured to play an important role in the production of seed nuclei in the alternate \textit{r}-process paths involving light neutron rich nuclei very near the $\beta$-stability line, and as such, it is important to know its ground state properties and structure to calculate rates of the reactions it might be involved in, in the stellar plasma. Found in the region of `island of inversion', its ground state might not be in agreement with normal shell model predictions.
\item[Purpose]
The aim of this paper is to study the elastic Coulomb breakup of $^{34}$Na on $^{208}$Pb to give us a core of $^{33}$Na with a neutron and in the process we try and investigate the one neutron separation energy and the ground state configuration of $^{34}$Na.
\item[Method]
A fully quantum mechanical Coulomb breakup theory within the architecture of post-form finite range distorted wave Born approximation extended to include the effects of deformation is used to research the elastic Coulomb breakup of $^{34}$Na on $^{208}$Pb at 100 MeV/u. The triple differential cross-section calculated for the breakup is integrated over the desired components to find the total cross-section, momentum and angular distributions as well as the average momenta, along with the energy-angular distributions.
\item[Results]
The total one neutron removal cross-section is calculated to test the possible ground state configurations of $^{34}$Na. The average momentum results along with energy-angular calculations indicate $^{34}$Na to have a halo structure. The parallel momentum distributions with narrow full widths at half maxima signify the same.

\item[Conclusion]
We have attempted to analyse the possible ground state configurations of $^{34}$Na and in congruity with the patterns in the `island of inversion' conclude that even without deformation, $^{34}$Na should be a neutron halo with a predominant contribution to its ground state most probably coming from $^{33}$Na($3/2^{+}$) $\otimes$ \textit{ $2p_{3/2}\nu$} configuration. We also surmise that it would certainly be useful and rewarding to test our predictions with an experiment to put stricter limits on its ground state configuration and binding energy.
 
\end{description}
\end{abstract}




\maketitle

\section {Introduction}

Exotic nuclei are a result of movement away from the valley of stability towards the drip line regions, where the increase in neutron (or proton) excess and low binding energy changes the nuclear structure, and leads to unconventional properties for the nuclei concerned.
Some of these exotic nuclei exhibit special character and are known as halo nuclei \cite{Khalili}. Nuclear halos are, essentially, a threshold effect occurring because of the presence of a bound state near the continuum in the energy spectrum \cite{Hansen} and as expected, neutron halos are more pronounced than proton halos owing to the large Coulomb barrier in the latter case. Characterized by a large spatial and hence, a small momentum distribution, they reveal interesting aspects about two- and three-body (Borromean) systems \cite{Zhukov}. Though they are highly unstable in most cases, the fact that they have non-negligible reaction rates in stellar plasma \cite{Rolf} has garnered the interest of experimentalists all over the world \cite{Tani, Naka, KAA, Kanungo}. Considerable interest has also been shown in exotic-halo nuclei ever since they first came to light \cite{Tani}. The theoretically explained models of the various reaction cycles, viz., the \textit{pp-}chains, CNO cycle, the \textit{r}-, and \textit{s}-processes, etc., led to the study of nuclei near the drip line in the medium mass region as they are also speculated to form connecting links within these chains which produce energy to power the stars \cite{Rolf,Iliadis,BBFH2,Bert}. In Ref. \cite{Hansen}, the authors explicitly state that it would be wise to carry out detailed research with exotic nuclei in the medium mass region.

Though these drip line halo nuclei have indeed been studied quite extensively for lower atomic masses \cite{Hansen, BanShy, Tani, NLV, RCEPJ, BayeDEA, PBan, BertSus}, the same, however, cannot yet be said about the medium mass region, where investigations have been comparatively fewer, but equally important nevertheless.
As has been customary with exotic nuclei, the studies in this region near the drip line \cite{Koba, Naka2, VT} have shown interesting characteristics in terms of their ground state configurations. Ground state (g.s.) configurations are important as their knowledge can lead one to the information about the isospin dependence of effective nuclear interaction, variation in the traditional shell structure or vanishing of the shell gaps, because even the `magic numbers' valid for the valley of stability get modified away from it \cite{Bastin, Nature, Moto, Iwasaki, 24O}. Normally, the nucleons inside the nucleus are expected to fill up according to the conventional shell model, but near the drip line the usual shell gaps break. Conventionally, one would expect the domination of the $\textit{f}_{7/2}$ orbit to form the ground state in nuclei in the vicinity of \textit{N} = 20 - 28. This would mean a large centrifugal barrier with \textit{l} = 3. But this does not favour the formation of halos as they are characterized by small angular momentum values (\textit{l} = 0, 1) to limit the effect of centrifugal barrier \cite{Rotival}. Nevertheless, in the `island of inversion' \cite{War}, rapid changes in the shell structure lead to configuration mixing due to the $\nu(sd)^{-2}(fp)^{2}$ intruder configurations, with $\nu$ representing the neutron. This configuration reversal results in a deformation in the nuclei away from the valley of stability. Deformation could be a factor for heavier nuclei surviving within the neutron drip line in this region \cite{Hama2012}.

Lying close to the drip line in this region, $^{34}$Na is such a nucleus. The progress in experimental technology over the years has made it possible to show that the one neutron separation energy ($S_{n}$) for $^{34}$Na is (0.17 $\pm$ 0.50) MeV \cite{Gaudefroy}, whereas the National Nuclear Data Centre (NNDC) database \cite{NNDC} shows it to be $\simeq$ (0.80 $\pm$ 0.008) MeV. Evidently, the uncertainty in the value cannot be ignored. Besides, its ground state spin-parity and shape (spherical or deformed, and oblate or prolate if the latter) have also not yet been established. If one goes by the recently observed trends around the mass region of \textit{N} = 20 - 28, for example in certain isotopes of Ne, Mg or Al, there is strong possibility that the ground state of $^{34}$Na is \textit{p}- or \textit{f}-wave dominant \cite{29Ne,31Ne,37Mg,35Mg,Al}. This knowledge is important because very weakly bound deformed nuclei in the island of inversion with significant $\textit{p}_{3/2}$ contribution could generate halos and $^{34}$Na is a strong candidate for both. Besides, the separation energy of a nucleus can also be used to calculate its matter radius \cite{FS} which can then be used to determine its shape. Further, $^{34}$Na is expected to be an innate fragment in the alternate \textit{r}-process paths where light to medium mass neutron rich nuclei play an important role either as or in the production of seed nuclei \cite{Tera}. $^{35}$Na is supposedly the most abundant isotope of Na near the neutron drip line and as such, binding energies of $^{34}$Na and $^{35}$Na are important to find their reaction rates for their production and consumption in stellar environments.

In this paper we use the method of Coulomb dissociation (CD) under the aegis of the post-form finite range distorted wave Born approximation (FRDWBA) theory \cite{37Mg, RC1} to evaluate different reaction observables for $^{34}$Na. CD, in principle, is the absorption of electromagnetic radiation by a projectile while it moves in the varying Coulomb field of a stable, heavy target and then fragmentizes into a core with one (or more) separate nucleon(s) \cite{Naka}. In this contribution, we apply the FRDWBA theory to calculate various reaction observables like total cross-sections, relative energy spectra, momentum and angular distributions, etc., and investigate the possible allowed ground state configurations for $^{34}$Na along with its binding energy. The post-form of the theory is fully quantum mechanical and includes the non-resonant continuum up to all orders. The only input that it requires is the ground state wavefunction of the projectile \cite{RC1}.

In the next section we present our formalism, while in Section \ref{sec:3} we show the results and discussion for the various quantities calculated for the reaction considered. Section \ref{sec:4} highlights the conclusions.


\section{Formalism}
\label{sec:2}
We contemplate that a beam of $^{34}$Na at 100 MeV/u impinges on a $^{208}$Pb target, and under its strong Coulomb influence, breaks up into $^{33}$Na and a neutron, i.e.,\\

$^{34}$Na + $^{208}$Pb $\longrightarrow$ $^{33}$Na + n + $^{208}$Pb.\\

\begin{figure}[h]
\centering
\includegraphics[height=8.3cm, clip,width=0.65\textwidth]{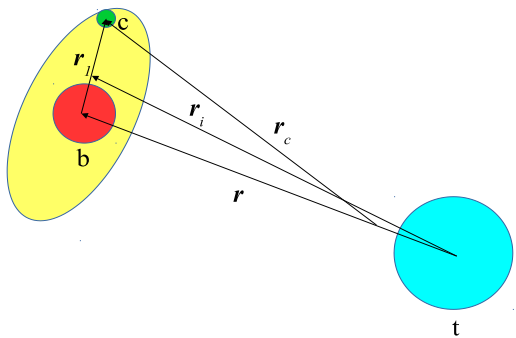}
\caption{\label{fig: 1} (Colour online) The three-body Jacobi coordinate system with a deformed projectile.}
\end{figure}

Considering the target to remain in its ground state (elastic breakup) and using the finite range distorted wave Born approximation theory, we first calculate the triple differential cross section and then integrate it to find different reaction observables for the breakup mentioned above. The coordinate system chosen is the Jacobi coordinate system shown in Fig.~\ref{fig: 1}.

It is trivial to verify that the position vectors \textbf{r$_{1}$}, \textbf{r$_{i}$}, \textbf{r$_{c}$} and \textbf{r} satisfy:
\begin{eqnarray}
{\bf{r}}={\bf{r}}_{i}-\alpha{\bf{r}}_{1} ;\hspace{0.2in} {\bf{r}}_{c}=\gamma{\bf{r}}_{1}+\delta{\bf{r}}_{i}\label{a1}.\nonumber    
\end{eqnarray}
with $\alpha, \delta$ and $\gamma$ being the mass factors given by:
\begin{eqnarray}
\alpha =\frac{{m_{c}}}{m_{c}+m_{b}};\hspace{0.2in}  \delta =\frac{m_{t}}{m_{b}+m_{t}};\hspace{0.2in} \gamma=(1-\alpha\delta),\label{a2}\nonumber
\end{eqnarray}
where, \textit{m$_{t}$}, \textit{m$_{b}$} and \textit{m$_{c}$} are the masses of the target \textit{t} ($^{208}$Pb), core \textit{b} ($^{33}$Na) and the valence nucleon \textit{c} (neutron), respectively.

Generically, the triple differential cross-section is defined as,
\begin{eqnarray}
\frac{d^3\sigma}{dE_{b}d\Omega_{b}d\Omega_{c}} = \frac{2\pi}{\hbar v_{at}}\rho_{(phase)}\frac{1}{\hat{j_{a}^{2}}}\sum_{\mu_{a}\mu_{b}\mu_{c}}|T^{(+)}_{fi}|^{2},\label{a4.1}
\end{eqnarray}
with \textit{a} being the $^{34}$Na projectile; $E_{b}$ is the energy of fragment \textit{b}, $\Omega$'s are the solid angles corresponding to fragments \textit{b} and \textit{c}, $v_{at}$ is the \textit{a-t} relative velocity in the initial channel, $j_{a}$ is the total spin of the projectile, $\mu$'s are the projections of the angular momenta of the corresponding particle and $\rho_{(phase)}$ is the three-body phase space factor \cite{Fuchs}. The transition matrix for post form in the FRDWBA is given by:
\begin{eqnarray}
T_{fi}^{(+)} = \int\int\int d\zeta d\textbf{r}_{1}d\textbf{r}_{i}\chi_{b}^{(-)*}(\textbf{q}_{b},\textbf{r})\phi_{b}^{*}(\zeta_{b})
\chi_{c}^{(-)*}(\textbf{q}_{c},\textbf{r}_{c})\phi_{c}^{*}(\zeta_{c}) V_{bc}(\textbf{r}_{1})
\phi_{a}^{lm}(\zeta_{a},\textbf{r}_{1})\chi_{a}^{(+)}(\textbf{q}_{a},\textbf{r}_{i}) \label{a4.2}
\end{eqnarray}
In Eq. (\ref{a4.2}), the $\chi$'s represent the pure Coulomb distorted waves with incoming [( - )] and outgoing [(+)] wave boundary conditions and $\textbf{q}_{a}, \textbf{q}_{b}$ and $\textbf{q}_{c}$ are the Jacobi wave vectors corresponding to $\textbf{r}_{1}, \textbf{r}$ and $\textbf{r}_{c}$, respectively. $\phi_{a}^{lm}(\zeta_{a}, \textbf{r}_{1})$ is the ground state wavefunction of \textit{a} having angular momentum \textit{l} and projection \textit{m} and internal coordinate $\zeta_{a}$. It includes the radial and the angular part of the projectile wavefunction ($\phi_{a}^{lm} = u_{l}(r_{1})Y^{m}_{l}(\hat{\textbf{r}}_{1})$). 

The \textit{T} - matrix, after integration over the internal coordinates ($\zeta$'s), yields,

\begin{eqnarray}
T_{fi}^{(+)} = \sum_{lmj\mu}\left\langle lmj_{c}\mu_{c}|j\mu\right\rangle \left\langle j_{b}\mu_{b}j\mu|j_{a}\mu_{a}\right\rangle i^{l}\hat{l}\beta_{lm}, \nonumber
\end{eqnarray}
such that 
\begin{equation}
|T^{(+)}_{fi}|^{2} = |\hat{l}\beta_{lm}|^{2}.\label{a4.3}
\end{equation}

where, $\hat{l} = \sqrt{(2l + 1)}$, and we define the reduced transition amplitude, $\beta_{lm}$, as:
\begin{eqnarray}
\hat{l}\beta_{lm} = \int\int d\textbf{r}_{1}d\textbf{r}_{i}\chi_{b}^{(-)*}(\textbf{q}_{b},\textbf{r})
\chi_{c}^{(-)*}(\textbf{q}_{c},\textbf{r}_{c})
 V_{bc}(\textbf{r}_{1})
\phi_{a}^{lm}(\textbf{r}_{1})\chi_{a}^{(+)}(\textbf{q}_{a},\textbf{r}_{i})\label{a4.4}
\end{eqnarray}

In case of particle \textit{c} being a neutron (case in point), $\chi_{c}^{(-)*}(\textbf{q}_{c},\textbf{r}_{c})$ is replaced by a plane wave ($e^{-i\textbf{q}_{c}.\textbf{r}_{c}}$) since then there is no Coulomb interaction between \textit{c} and \textit{t}.

We emphasize that the deformation enters our theory via the quadrupole-deformed potential \textit{$V_{bc}$} in Eqs. (\ref{a4.2}) and (\ref{a4.4}) \cite{Hama},

\begin{eqnarray}
V_{bc}(\textbf{r}_{1}) = V_{s}(r_{1}) - \beta_{2}V_{ws}
R\left[  \frac{d{g(r_{1})}}{dr_{1}} \right] Y_{2}^{0}(\hat{\textbf{r}}_{1}), \label{a4.5}
\end{eqnarray}
where \textit{V$_{ws}$} is the Woods-Saxon potential depth, $\beta_{2}$ is the quadrupole deformation parameter and \textit{g}(r$_{1}$) = $\left[{1 + exp(\frac{r_{1} - R}{a})}\right]^{-1}.$

Here, \textit{V$_{s}$}(r$_{1}$) $=\textit{V$_{ws}$}\textit{g}(r_{1})$ \textendash with radius $R = r_{0}A^{1/3}$, $r_{0}$ and  \textit{a} being the radius and diffuseness parameters, respectively, while \textit{A} is the mass number of the projectile \textendash represents the spherical part of the Woods-Saxon potential. In effect, Eq. (\ref{a4.5}) is the pruned Taylor expansion of the potential $V_{bc}(\textbf{r}_{1})$, in which we have neglected the spin-orbit term in defining it axially symmetrically. The values of $r_{0}$ and \textit{a} have been taken throughout to be 1.24 fm and 0.62 fm, respectively. 

The radial wavefunctions ($u_{lm}(r_{1})$) should be obtained from the coupled equations which result due to the presence of deformation in the axially symmetric potential \cite{Bohr}:

\begin{eqnarray}
\left\{ \frac{d^2}{dr_1^2} -\frac{l(l+1)}{r_1^2} +\frac{2\mu}{\hbar^2}
[E - V_{s}(r_1)]\right\} u_{l m}(r_1) = 
\frac{2\mu}{\hbar^2}\sum_{l^\prime} \langle Y_{l}^m(\hat {\bf r}_1)|- \beta_2 k(r_1) 
Y^{0}_{2}(\hat {\bf r}_1)|Y_{l^\prime}^m(\hat {\bf r}_1)\rangle u_{l^\prime m}(r_1).\nonumber\\
~
\end{eqnarray}
For a given \textit{l} value, these will have an admixture of other \textit{l} components of matching parity as well.
This would mean that positive parity levels like \textit{s}- states can have admixtures from other \textit{s}- and \textit{d}-, \textit{g}-, etc. orbitals while the negative parity levels like the \textit{p}- states can have mixing contributions from other \textit{p}- as well as \textit{f}-, \textit{h}-, levels, which renders the orbital angular momentum to no longer be conserved.
Nevertheless, it has been found that for weak components of higher \textit{l} states, the admixtured states will be ruled by pure states of lowest \textit{l} values \cite{Hama, Misu}. In fact, Ref. \cite{Hama} clearly verifies that for the case of very low binding energy of the valence neutron (tending to zero), lowest \textit{l} component dominates in the neutron orbits of a realistic deformed potential, and this is independent of the extent of deformation.

Hence, although we have calculated the radial wavefunction from a spherical Woods-Saxon potential for a single \textit{l}, because the binding energy of $^{34}$Na is very low, it is not too unreasonable to use it in conjunction with an axially symmetric quadrupole deformed Woods-Saxon potential for further calculations \cite{37Mg, 31Ne}.


Eq. (\ref{a4.4}) is a cumbersome integral which does not converge easily. We circumvent the problem by Taylor expanding $\chi_{b}^{(-)*}(\textbf{q}_{b},\textbf{r})$ about \textbf{r$_i$} and replacing the `del' operator by an effective local momentum approximation (LMA) \cite{Shyam, Ban} and then write,

\begin{eqnarray}
\chi_{b}^{(-)*}(\textbf{q}_{b},\textbf{r}) = e^{\alpha\nabla_{r_{i}}.\textbf{r}_{1}}\chi_{b}^{(-)*}(\textbf{q}_{b},\textbf{r}_{i})\nonumber\\ \xrightarrow{LMA}e^{i\alpha\textbf{K}.\textbf{r}_{1}}\chi_{b}^{(-)*}(\textbf{q}_{b},\textbf{r}_{i}), \label{a4.6}
\end{eqnarray}

\textbf{K} being the local momentum of the core \textit{b}, the magnitude of which is given by:

\begin{equation}
K = \sqrt{\dfrac{2m_{bt}}{\hbar^{2}}(E_{bt} - V(R))} \label{a4.7}
\end{equation}
where $m_{bt}$ is the reduced \textit{b-t} mass and $E_{bt}$ is the relative energy of the \textit{b-t} system. The Coulomb potential between \textit{b} and \textit{t} at a distance \textit{R} is represented by \textit{V(R)}.

This results in a factorization of the six-dimensional reduced transition amplitude of Eq. (\ref{a4.4}) into two parts: one, the dynamics part and two, the structure part:

\begin{eqnarray}
\hat{l}\beta_{lm} = \int d\textbf{r}_{i}e^{-i\delta\textbf{q}_{c}.\textbf{r}_{i}} \chi_{b}^{(-)*}(\textbf{q}_{b},\textbf{r}_{i})\chi_{a}^{(+)}(\textbf{q}_{a},\textbf{r}_{i}) 
\int d\textbf{r}_{1}e^{-i\textbf{W}.\textbf{r}_{1}}V_{bc}(\textbf{r}_{1})\phi_{a}^{lm}(\textbf{r}_{1}) \label{a4.8}
\end{eqnarray} 
In Eq. (\ref{a4.8}), the plane wave for the neutron is surrogated by $e^{-i\textbf{q}_{c}.(\gamma\textbf{r}_{1} + \delta\textbf{r}_{i})}$ using the defined relations for position vectors, and hence, $\textbf{W} = \gamma\textbf{q}_{c} - \alpha\textbf{K}$.
As the nomenclature suggests, the first integral in Eq. (\ref{a4.8}) controls the dynamics part in the breakup and can be expressed analytically in terms of the Bremsstrahlung integral \cite{Bremsstrahlung}. The second integral in Eq. (\ref{a4.8}) holds information about the structure of the nucleus under consideration and thus, about the effects that any deformation present in it might produce. Consequently, with changes in the shape of the nucleus, the dynamics part of the reduced transition amplitude remains unaffected and it is only the structure part that gets modified.

For further details on the formalism, one may refer to \cite{31Ne, RC1}.

The results are discussed in the following section.
\section{Results and discussions}
\label{sec:3}

In the present text, we have inspected one neutron removal from the \textit{$2s_{1/2}$}, \textit{$2p_{3/2}$} and \textit{$1f_{7/2}$} orbitals that can possibly contribute to $^{34}$Na ground state. The separation energy for $^{34}$Na is not very well defined and we have varied it from 0.01 MeV to 0.80 MeV, ultimately fixing a value of 0.17 MeV (as measured in Ref. \cite{Gaudefroy}) for further calculations.

\begin{figure}[htbp]
\centering
\includegraphics[trim={0 0 0 0},clip,width=8.3cm]{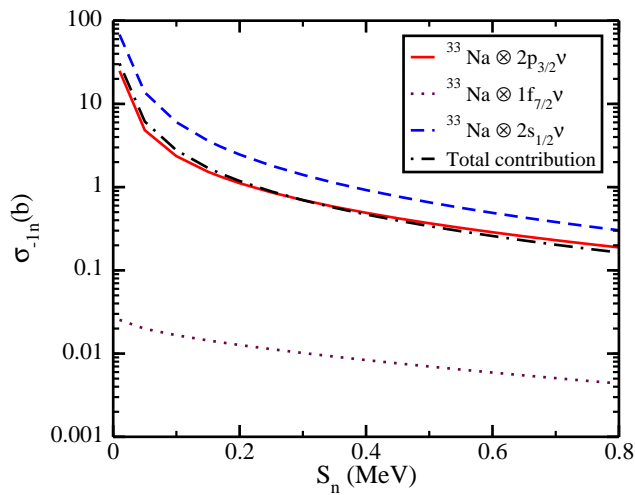}

\caption{\label{fig: 2} (Colour online) The total cross-section $(\sigma_{-1n})$ vs one neutron separation energy ($S_{n}$) for $^{34}$Na breaking on $^{208}$Pb at 100 MeV/u beam energy. The deformation parameter, $\beta_2$, has been set to 0. The solid line shows the cross-section for \textit{p}-wave neutron configuration while the dashed line shows the same from the \textit{s}-wave. The dotted line is for the \textit{f}-wave configuration whereas the dash-dotted line gives the effect of equal contribution from all the states taken together with spectroscopic factor (\textit{C$^{2}$S}) for each being equal to 0.33.}

\label{}
\end{figure}

In Fig. \ref{fig: 2}, we present the total one neutron removal cross-section $(\sigma_{-1n})$ for the elastic Coulomb breakup of $^{34}$Na on a $^{208}$Pb target at 100 MeV/u beam energy as a function of one neutron separation energy for different possible cases of ground state configurations ($^{33}$Na $\otimes$ \textit{$2s_{1/2}\nu$}, $^{33}$Na $\otimes$ \textit{$2p_{3/2}\nu$}, $^{33}$Na $\otimes$ \textit{$1f_{7/2}\nu$}) of $^{34}$Na. The $^{33}$Na core has a ground state spin-parity of $3/2^{+}$. The deformation parameter for $^{34}$Na is taken to be zero for the present case.
It can be clearly seen that the cross-section for the \textit{$1f_{7/2}$} (dotted line) neutron configuration is orders of magnitude lesser than the cross-sections for \textit{$2s_{1/2}$} (dashed line) or \textit{$2p_{3/2}$} (solid line) configurations. Keeping in mind that the ground state might be an admixture of more than one state, we have also shown in Fig. \ref{fig: 2} the total cross-section if the ground state is considered to be equally contributed (i.e., spectroscopic factor, \textit{C$^{2}$S}, is 0.33 for each state so that the total \textit{C$^{2}$S} is 1)\footnote{Spectroscopic factors are important quantities not only related to single-particle orbitals but are also vital in the calculations of thermonuclear reaction rates in astrophysics \cite{Wiescher, Descouvemont, Mukhamedzhanov}. In most cases, spectroscopic factors are deduced by matching a theoretical calculation with an experimental observable (see eg., Refs. \cite{37Mg, KobaPRL} for the recent case of $^{37}$Mg) and enter the analysis as multiplicative coefficients \cite{Wiescher, Descouvemont}.} by all the three states considered (the dash-dotted curve). The curve seems to merge with the \textit{p-} wave contribution. Evidently, the \textit{$1f_{7/2}$} contribution is almost negligible. Moreover, the pattern observed off late in the mass range of \textit{N} = 20 - 28 suggests that due to the `island of inversion', the probability that $^{34}$Na, in its ground state, has a dominant \textit{$\nu(2s_{1/2})$} state becomes remote again \cite{37Mg, 31Ne, 29Ne, 35Mg}. Thus, if there is any chance of $^{34}$Na having a mixed configuration, it is most likely to have a dominant \textit{$2p_{3/2}$} contribution to its ground state. A more conclusive evidence for the ground state of $^{34}$Na would be to calculate its relative energy spectra and see the effects of deformation on it.

Given the experimental data, our estimates could be used to predict the binding energy of the projectile.

\begin{figure}[htbp]
\centering
\includegraphics[trim={0 0 0 0},clip,width=8.3cm]{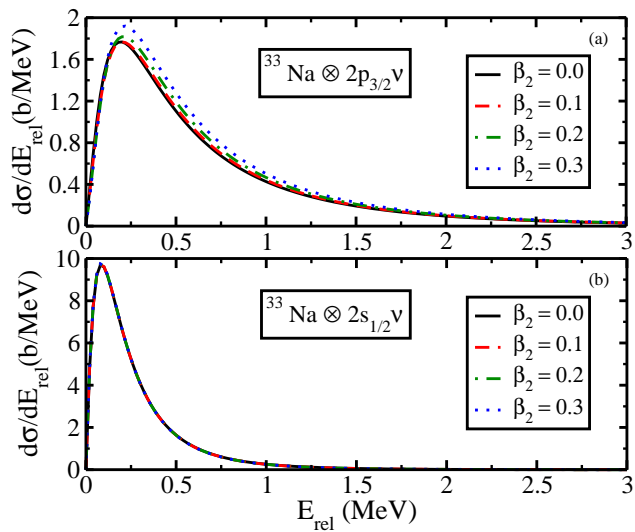}

\caption{\label{fig: 3} (Colour online) The relative energy spectra of $^{34}$Na breaking elastically on $^{208}$Pb at 100 MeV/u beam energy due to Coulomb dissociation. For a neutron separation energy of 0.17 MeV, the deformation parameter, $\beta_2$ is varied from 0.0 to 0.3. (a) The g.s. for $^{34}$Na is assumed to be formed from contribution of \textit{$2p_{3/2}$} valence neutron. (b) The same when the g.s. is formed by \textit{$2s_{1/2}$} contribution.}

\label{}
\end{figure}

The relative energy spectra was also calculated for different values of deformation parameter $\beta_{2}$ for a neutron separation energy of 0.17 MeV, where we consider two separate cases: (a) the valence neutron in \textit{$2p_{3/2}$} orbital forms the ground state for $^{34}$Na, and (b) the valence neutron is in \textit{$2s_{1/2}$}. The results are shown in Fig. \ref{fig: 3}.

As is clear from Fig. \ref{fig: 3}(a), the peak height for the cross-section increases with increasing the deformation of the nucleus. Nevertheless, the shape of each of the curves essentially remains the same. In Fig. \ref{fig: 3}(b), one can see that there is hardly any effect of deformation on the relative energy spectra. This is because the \textit{s}-wave configuration, in our calculations, does not provide any constraints on $\beta_{2}$ \cite{37Mg}. However, amplitude wise, the peak height in this case is much higher than that obtained when $^{34}$Na is assumed to have the neutron in \textit{ $2p_{3/2}$} orbital in its ground state. In this regard, experimental results for $^{34}$Na breaking up due to Coulomb dissociation would be of immense help in limiting the uncertainty in its ground state configuration as well as its binding energy. A superior knowledge of these quantities can help explain better the quadrupole deformation of the sodium isotope.

\begin{figure}[htbp]
\centering
\includegraphics[trim={0 0 0 0},clip,width=8.3cm]{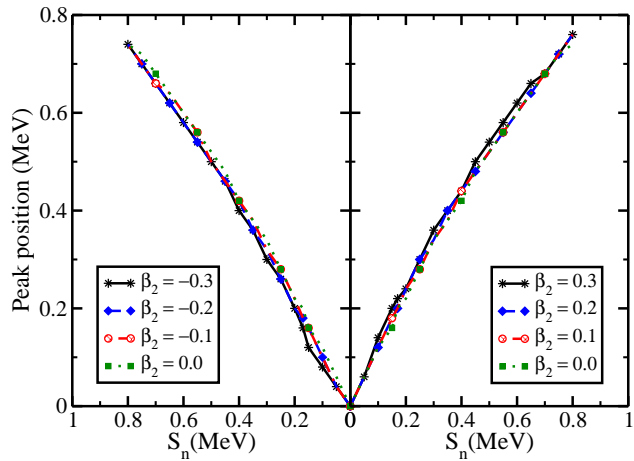}
\caption{\label{fig: 4} (Colour online) Peak position of relative energy spectra as a function of $S_{n}$ in the breakup of $^{34}$Na on $^{208}$Pb at 100 MeV/u for deformation parameter, $-0.3\leq\beta_{2}\leq0.0$ on the left, and $0.0\leq\beta_{2}\leq0.3$ on the right. The average slope of all the curves is 0.9365. The g.s. of $^{34}$Na is supposed to be formed by contribution of $\nu$(\textit{$2p_{3/2}$}) state. The lines are a guide to the eye.}

\label{}
\end{figure}

\begin{table}
\begin{center}

\caption{\label{T2} Slope of the curves obtained for different values of deformation parameter, $\beta_2$, in Fig. 4. The average slope comes out to be 0.9365.}
\vspace{0.50cm}
\begin{tabular}{ |c|c|c|c| } 
\hline\hline
 $\beta_2$ & Slope  \\
\hline
 -0.3 & 0.9515 \\
 -0.2 & 0.9318 \\
 -0.1 & 0.9370 \\
  0.0 & 0.9355 \\
  0.1 & 0.9356 \\
  0.2 & 0.9323 \\
  0.3 & 0.9320 \\

\hline\hline
\end{tabular}

\end{center}
\end{table}

As stated earlier, the limits on its binding energy are too vague and the vagueness about $^{34}$Na is accentuated by the fact that its shape is also not known (spherical or deformed). Since it lies in the `island of inversion', it is speculated to have a deformed structure. Fig. \ref{fig: 3} clearly shows the peak position of the relative energy spectra being affected more by quadrupole deformation parameter, $\beta_{2}$ for $^{33}$Na $\otimes$ \textit{$2p_{3/2}\nu$} g.s. configuration than for $^{33}$Na $\otimes$ \textit{$2s_{1/2}\nu$}. We believe that simple scaling laws can help reduce these uncertainties.

For loosely bound nuclei, it is known that the dipole strength, $B(E1)$, is inversely related to the binding energy of the projectile, $S_{n}$ \cite{NLV, TB, BCH}, and the total dipole Coulomb breakup cross-section is proportional to the electric dipole reduced probability. The inverse-law dependence of our total cross-section with respect to the binding energy (in Fig. \ref{fig: 2}) points to the possibility that $^{34}$Na could be dominated by an \textit{E1} transition \cite{NLV} because even for nuclei in the vicinity of the drip line, the chief contribution to the breakup process is predominantly by dipole dissociation. The higher order multipole contributions are negligible \cite{RCEPJ}. Thus, knowing the dipole strength functions from the experiments can also give us an idea about the binding energy of $^{34}$Na.

In Ref. \cite{RCEPJ}, it is argued that scaling is indeed valid for our fully quantum mechanical approach and the authors build on the fact that the peak position of the relative energy spectra is directly proportional to the binding energy of the nucleus. In Fig. \ref{fig: 4}, we show the peak position of the relative energy spectra as a function of one-neutron separation energy. It is observed from the graph that the dependence is almost linear and more importantly, the pattern appears to be independent of the sign of the quadrupole deformation. Table \ref{T2} displays the slope of the linear fitted curves in Fig. \ref{fig: 4}, and it is evident that their values are almost same, varying only in the third place of decimal. The value of average slope of the curves comes out to be 0.9365. Given the peak position of the experimental relative energy spectra, the linear dependence of peak position on separation energy can be used to get an investigatory estimate about the binding energy of the projectile, which is known but associated with a large uncertainty \cite{Gaudefroy}.

\begin{figure}[htbp]
\centering
\includegraphics[trim={0 0 0 0},clip,width=8.3cm]{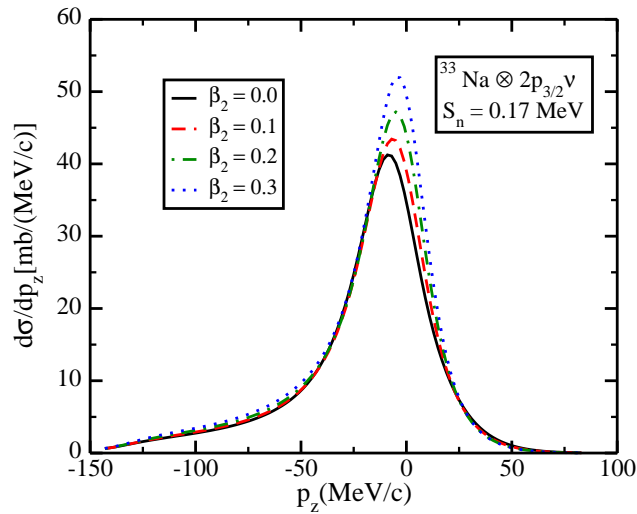}

\caption{\label{fig: 5} (Colour online) The parallel momentum distribution of $^{33}$Na for $^{34}$Na breaking elastically on $^{208}$Pb at 100 MeV/u beam energy. $^{33}$Na $\otimes$ \textit{ $2p_{3/2}\nu$} is assumed to form the ground state of $^{34}$Na for a neutron separation energy of 0.17 MeV. The solid, dashed, dash-dotted and dotted lines correspond to a quadrupole deformation value of 0.0, 0.1, 0.2 and 0.3, respectively.}

\label{}
\end{figure}

\begin{table}[h]
\begin{center}

\caption{\label{T3} Full width at half maximum of the parallel momentum distribution of $^{33}$Na for different values of deformation parameter, $\beta_2$, yielded when $^{34}$Na breaks up elastically on $^{208}$Pb at 100 MeV/u due to Coulomb forces. The projectile ground state corresponds to $^{33}$Na $\otimes$ \textit{ $2p_{3/2}\nu$} configuration with one-neutron separation energy, $S_{n}$ = 0.17 MeV.}
\vspace{0.50cm}
\begin{tabular}{ |c|c|c|c| } 
\hline\hline
$S_{n}$ (MeV) & $\beta_2$ & FWHM (MeV/c)  \\
\hline
 
 0.17 & 0.0 & 36.82 \\
      & 0.1 & 35.88 \\
      & 0.2 & 34.76 \\
      & 0.3 & 33.83 \\

\hline\hline
\end{tabular}

\end{center}
\end{table}

\begin{figure}[htbp]
\centering
\includegraphics[trim={0 0 0 0},clip,width=8.3cm]{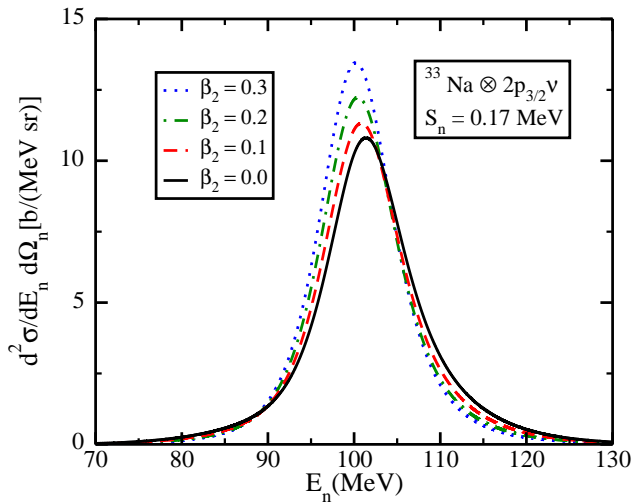}

\caption{\label{fig: 6} (Colour online) The neutron energy-angular distribution for $^{34}$Na breaking on $^{208}$Pb at 100 MeV/u for separation energy of 0.17 MeV. The ground state of $^{34}$Na is assumed to be given by $^{33}$Na $\otimes$ \textit{ $2p_{3/2}\nu$}. The solid, dashed, dash-dotted and dotted curves correspond to a quadrupole deformation ($\beta_2$) value of 0.0, 0.1, 0.2 and 0.3, respectively. The neutron angle, $\theta_{n} = 1^{\circ}$.}

\label{}
\end{figure}

Fig. \ref{fig: 5} brings forth the longitudinal or the parallel momentum distribution (PMD) calculations (for parallel momentum range calculated using, $p_{b} = (\sqrt{2m_{b}E_{b}} \pm 150)$ MeV/c) for the $^{33}$Na fragment obtained in the elastic Coulomb breakup of $^{34}$Na on $^{208}$Pb at 100 MeV/u beam energy. The investigations are done for $^{33}$Na$(3/2^{+})$ $\otimes$ \textit{$2p_{3/2}\nu$} configuration as the ground state of $^{34}$Na. Unlike the transverse momentum distributions, PMDs are relatively insensitive to the nuclear interaction and beam energy controlling the breakup \cite{Greiner, Kelley, Orr}, and if narrow, reflect an extended spatial distribution via Heisenberg's uncertainty principle.

We observe narrow parallel momentum distributions which point to broader spatial distributions. It is evident from the figure as well as from the values in Table \ref{T3} that deformation in the nucleus makes the full width at half maxima (FWHM) of the distributions narrower (the peak height rises with increase in deformation). But the FWHM value even for zero quadrupole deformation is $<$44 MeV/c (the half-maximum width for established halos like $^{11}$Be and $^{19}$C \cite{Ban, RC1}), suggesting that  $^{34}$Na is indeed halo, even in the absence of any deformation and if at all, its halo character only increases with increasing $\beta_{2}$.

The neutron energy-angular distribution for our breakup reaction is displayed in Fig. \ref{fig: 6} for which the neutron angle, $\theta_{n}$, was fixed at $1^{\circ}$ to correspond to the case of near or below the grazing angles in the forward direction. Similar to the case of the momentum distributions, the amplitude increases with increase in the value of $\beta_2$. The increase in amplitude confirms the narrowing of the FWHM of the distribution peaks obtained. The effects of deformation are clearly seen to be more near the peaks of the distributions. Irrelevant of deformation, the peak position in the Fig. \ref{fig: 6} almost coincides with the beam energy. This is an indication that there should be no post-acceleration effects for the charged fragment \cite{RC1}. No post-acceleration effect is another feature in the breakup of halo nuclei \cite{RShyam}.

\begin{figure}[htbp]
\centering
\includegraphics[trim={0 0 0 0},clip,width=8.3cm]{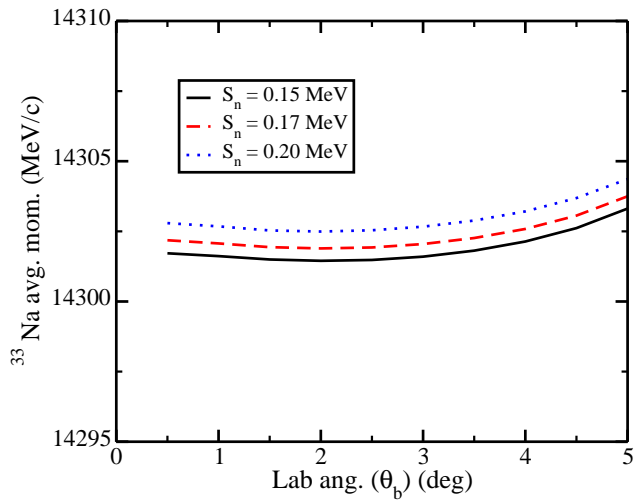}

\caption{\label{fig: 7} (Colour online) Average momentum of the $^{33}$Na fragment in the breakup of $^{34}$Na on $^{208}$Pb at 100 MeV/u as function of its detection angle.}

\label{}
\end{figure}

\begin{figure}[htbp]
\centering
\includegraphics[trim={0 0 0 0cm},clip,width=8.3cm]{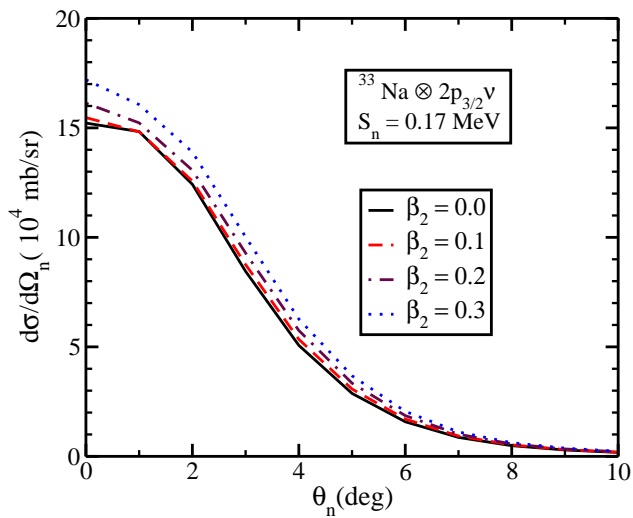}

\caption{\label{fig: 8} (Colour online) The angular distribution for $^{34}$Na breaking on $^{208}$Pb at 100 MeV/u beam energy to form $^{33}$Na and a neutron. The deformation parameter, $\beta_2$ is varied from 0.0 to 0.3 for a neutron separation energy of 0.17 MeV. The g.s. of $^{34}$Na is supposed to be formed by \textit{$2p_{3/2}$} configuration of the valence neutron.}

\label{}
\end{figure}

Post-acceleration, from semiclassical arguments \cite{Baur}, should come to light with increase of average momentum \cite{RC1} of the charged fragment with the scattering angle. To check that, we calculated the average momentum of $^{33}$Na and plotted it with respect to its scattering angle in Fig. \ref{fig: 7}. According to the semiclassical point of view, Coulomb repulsion effects on the charged fragment increase with decreasing of the impact parameter. This is expected because classically, as the scattering angle, $\theta_b$, increases, impact parameter decreases, resulting in an increased Coulomb repulsion from the target nucleus. However, Fig. \ref{fig: 7} depicts that the average momentum of the $^{33}$Na fragment is almost saturated for the very forward angles. Even as the angle increases, the percentage change in the average momentum is extremely small, indicating that the breakup distance is still large so that the $^{33}$Na fragment is not accelerated on its way out. Hence, post-acceleration may be ruled out. The variation of average momentum with binding energy of the valence neutron only results in an increase in its absolute value while essentially reproducing similar curves.

The neutron angular distribution for the breakup is seen in Fig. \ref{fig: 8}. One notices that the differential cross-section falls steeply as the neutron angle increases in the forward direction. The effect of deformation is seen to be prominent at very forward angles. The constricted angular distributions for Coulomb breakup of $^{34}$Na below grazing angles can also be used to study its halo effect \cite{Esbensen}. Narrow angular distributions are in agreement with smaller widths of parallel momentum distribution \cite{31Ne} which further boosts the impression that $^{34}$Na could be a halo nucleus.



\section{Conclusions}
\label{sec:4}

In the present work, we try to analyse the halo character of $^{34}$Na while calculating different reaction observables if it undergoes elastic Coulomb breakup when bombarded on $^{208}$Pb at 100 MeV/u and in doing so, try and get a better understanding of its one neutron separation energy and ground state configuration. We apply the Coulomb dissociation method under the patronage of the finite range distorted wave Born approximation extended to include the effects of deformation in an approximate way. This theory requires only the ground or the bound state wavefunction of the projectile and includes the entire non-resonant continuum. We are able to factorize the reduced transition amplitude into a structure part and a dynamics part. The theory, with and without the inclusion of deformation, has been used in the past to study various nuclei as well as their radiative capture reactions \cite{31Ne, 37Mg, Neelam, PbRcRs, 15C}. 

Our results, combined with the patterns detected in the medium mass region of `island of inversion' with \textit{N} = 20 - 28, augment the speculation that $^{34}$Na has a dominant \textit{p}-state configuration, which also support the suggestions by Refs. \cite{FS, Gaudefroy}. Moreover, there are chances that more often than not we would encounter a halo nucleus in this region. The halo structure of $^{34}$Na is manifested via the narrow longitudinal momentum distributions with decreasing FWHM values for increasing values of deformation, and is further corroborated by its forward peaked angular distributions. The average momentum calculations along with the energy-angular distributions show that the breakup of $^{34}$Na on $^{208}$Pb was free from post-acceleration effects.

The one neutron removal cross-section observed can be equated with future experimental data to put more stringent bounds on the one neutron removal energy value and the ground state configuration (whether it is an \textit{s-, p-} or an \textit{f}- wave). In fact, these limits can further be improved using the relative energy spectra results whose peaks varied with defromation. The peak position of relative energy spectra can be used in future for scaling purposes. At this beam energy range ($\sim$ a few hundred MeV/u), the final channel fragments are usually easier to detect as they are ejected with higher velocities. This facilitates measuring Coulomb dissociation observables like the angular distributions or the relative energy spectra, using which one would be able to put constraints on spectroscopic factors.

The relative energy spectra can also be used with the principle of detailed balance to find the relevant $^{33}$Na$(n,\gamma)^{34}$Na radiative capture cross-section to different possible low-lying states of $^{34}$Na. This would be important to get an idea about $^{35}$Na being the most abundant isotope of sodium near the neutron drip line \cite{Tera}.

We strongly encourage experiments to affirm the predictions and put firmer limits on the values of binding energy and ground state configuration for $^{34}$Na.

\section*{Acknowledgment}
This text results from research supported by the Department of Science and Technology, Govt. of India, (SR/S2/HEP-040/2012). Support from MHRD grant, Govt. of India, to [GS] is gratefully acknowledged. [S] is supported by the U.S. NSF Grant No. PHY-1415656 and the U.S. DOE Grant No. DE-FG02-08ER41533.




\end{document}